\documentclass[twocolumn, prx, superscriptaddress,notitlepage]{revtex4-2}
\usepackage{graphicx}
\usepackage{dcolumn}
\usepackage{bm}
\usepackage[usenames,dvipsnames]{color}
\usepackage[most]{tcolorbox}
\usepackage{multirow}
\usepackage{gensymb}
\usepackage[normalem]{ulem}
\usepackage{CJK}
\usepackage{comment}
\usepackage[colorlinks, linkcolor=blue,anchorcolor=blue,citecolor=blue,urlcolor=blue]{hyperref}
\usepackage{amssymb}
\usepackage{pifont}
\usepackage{physics}
\usepackage{natbib}
\usepackage{xcolor}
\usepackage{siunitx}

\usepackage{color,soul}

\begin{document}
\begin{CJK*}{UTF8}{}
\title{Programmable few-atom Bragg scattering and ground-state cooling in a cavity}

\author{Guoqing Wang}
\thanks{These authors contributed equally.}
\affiliation{
   MIT-Harvard Center for Ultracold Atoms and Research Laboratory of Electronics, Massachusetts Institute of Technology, Cambridge, MA 02139, USA}
\affiliation{
   Department of Physics, Massachusetts Institute of Technology, Cambridge, MA 02139, USA}

\author{David C. Spierings}
\thanks{These authors contributed equally.}
\affiliation{
   MIT-Harvard Center for Ultracold Atoms and Research Laboratory of Electronics, Massachusetts Institute of Technology, Cambridge, MA 02139, USA}
\affiliation{
   Department of Physics, Massachusetts Institute of Technology, Cambridge, MA 02139, USA}

\author{Matthew L. Peters}
\thanks{These authors contributed equally.}
\affiliation{
   MIT-Harvard Center for Ultracold Atoms and Research Laboratory of Electronics, Massachusetts Institute of Technology, Cambridge, MA 02139, USA}
\affiliation{
   Department of Physics, Massachusetts Institute of Technology, Cambridge, MA 02139, USA}

\author{Meng-Wei Chen}
\affiliation{
   MIT-Harvard Center for Ultracold Atoms and Research Laboratory of Electronics, Massachusetts Institute of Technology, Cambridge, MA 02139, USA}
\affiliation{
   Department of Physics, Massachusetts Institute of Technology, Cambridge, MA 02139, USA}

\author{Uro\v{s} Deli\'{c}}
\affiliation{Vienna Center for Quantum Science and Technology, Atominstitut, TU Wien, A-1020 Vienna, Austria}
\affiliation{
   University of Vienna, Faculty of Physics, Vienna Center for Quantum Science and Technology, A-1090 Vienna, Austria}

\author{Vladan Vuleti\'c}
\email[]{vuletic@mit.edu}
\affiliation{
   MIT-Harvard Center for Ultracold Atoms and Research Laboratory of Electronics, Massachusetts Institute of Technology, Cambridge, MA 02139, USA}
\affiliation{
   Department of Physics, Massachusetts Institute of Technology, Cambridge, MA 02139, USA}

\begin{abstract}
By integrating tweezer arrays with a high-cooperativity ring cavity with chiral atom-cavity coupling, we demonstrate highly directional Bragg scattering from a programmable number of atoms. Through accurate control of the interatomic distance, we observe a narrowing-down of the Bragg peak as we increase the atom number one by one.
The observed high-contrast Bragg interference is enabled by cavity sideband cooling of both the radial and axial motions to near the ground state with phonon occupation numbers below 0.17 and 3.4, respectively.  This new platform that integrates strong and controlled atom-light coupling into atomic arrays enables applications from programmable quantum optics to quantum metrology and computation.  
\end{abstract}

\maketitle

\end{CJK*}	

Arrays of individual atoms trapped in optical tweezers are a promising platform for quantum computation~\cite{bluvstein_logical_2024,bluvstein_quantum_2022}, simulation~\cite{bernien_probing_2017,ebadi_quantum_2021,semeghini_probing_2021}, and metrology~\cite{finkelstein_universal_2024,cao_multi-qubit_2024} in large part due to the ability to configure the geometric arrangement of the atoms, and control them individually or in groups. Recent progress in this field includes increasing array sizes to thousands of atoms~\cite{manetsch_tweezer_2024,norcia_iterative_2024}, demonstrating logical quantum operations~\cite{bluvstein_logical_2024,reichardt_logical_2024}, and achieving two-qubit gate fidelity exceeding 99.5\%~\cite{evered_high-fidelity_2023,finkelstein_universal_2024}. These breakthroughs are rooted in the high level of control over light-matter interactions, often using cavity quantum electrodynamics (cavity QED) platforms to achieve experimental control at the level of a single photon and a single atom~\cite{Rempe2015_review, Ritch2021_review}. The integration of tweezer arrays with optical cavities has already yielded a number of advances, including deterministic control of atom number~\cite{shadmany_cavity_2024,menon_integrated_2024,liu_realization_2023,zhang_cavity_2024}, fast atomic-state readout~\cite{deist_mid-circuit_2022,peters_cavity-enabled_2024}, classical error correction~\cite{hu_site-selective_2024}, and entanglement generation with error detection~\cite{grinkemeyer_error-detected_2024}.

Optical cavities are uniquely suited to enhance collective light-scattering, which can depend strongly on the spatial arrangement of the atoms through interference effects. In free space, Bragg scattering has been observed in one-~\cite{slama_phase-sensitive_2005,schilke_photonic_2011,sorensen_coherent_2016,corzo_large_2016}, two-~\cite{weitenberg_coherent_2011}, and three-dimensional optical lattices~\cite{birkl_bragg_1995,weidemuller_bragg_1995}. Collective scattering in free space has been used to probe geometric structure and atomic motion~\cite{birkl_bragg_1995,weidemuller_bragg_1995,westbrook_study_1997,raithel_cooling_1997}, magnetic structure~\cite{weitenberg_coherent_2011,hart_observation_2015}, phase transitions~\cite{miyake_bragg_2011}, and to control the propagation direction of light~\cite{rui_subradiant_2020,srakaew_subwavelength_2023}. 
Recently, tweezer arrays in Fabry-P\'erot cavities have been used to study collective light-scattering in a more flexible manner, enabling observation of super- and sub-radiance~\cite{yan_superradiant_2023} as well as self-organization in the standing-wave modes of the two-mirror cavity~\cite{ho_optomechanical_2024}, which builds upon previous work using bulk gases~\cite{Black_2003,Esslinger_2010,Lev_2018}.

Another key requirement to realize robust control of atoms in tweezers is that the atoms must remain cold, ideally near the vibrational ground state, as finite temperature can limit the fidelity of the light-atom interaction and reduce the contrast of interatomic interference effects. Standard free-space laser cooling methods for atoms in tweezers rely on the internal atomic levels, such as polarization gradient cooling~\cite{bluvstein_quantum_2022}, Raman sideband cooling~\cite{kaufman_cooling_2012,thompson_coherence_2013-1,lester_raman_2014,spence_preparation_2022,tian_resolved_2024}, and $\Lambda$-enhanced gray molasses~\cite{blodgett_imaging_2023,angonga_gray_2022}. On the other hand, cavity cooling techniques are limited by the spectral properties of the cavity rather than the atom, and therefore represent an interesting alternative, e.g., to perform cooling at an arbitrary detuning from atomic resonance, or when closed transitions or an appropriate atomic multi-level structure are not available~\cite{Vuletic2000,vuletic_three-dimensional_2001}.

In this work, we demonstrate two-dimensional ground-state cavity cooling of atoms in a tweezer array to enable the further study of few-atom Bragg scattering. By tuning the array's geometric structure on a sub-wavelength scale, we observe the emergence of Bragg's condition and the narrowing of the Bragg peak in each cavity mode as we increase the number of atoms. Using the incident beam angle and the array's geometry as control parameters, we observe tunable and directional light scattering. Additionally, inversion symmetry is broken in our twisted ring cavity, splitting the chiral degeneracy of the polarization degree of freedom of cavity light. As a result, when an atom in a magnetically sensitive state further breaks the time-reversal symmetry of the atom-light interaction, we realize chiral light scattering even at the single-atom level.

\begin{figure*}
    \includegraphics[width=\textwidth]{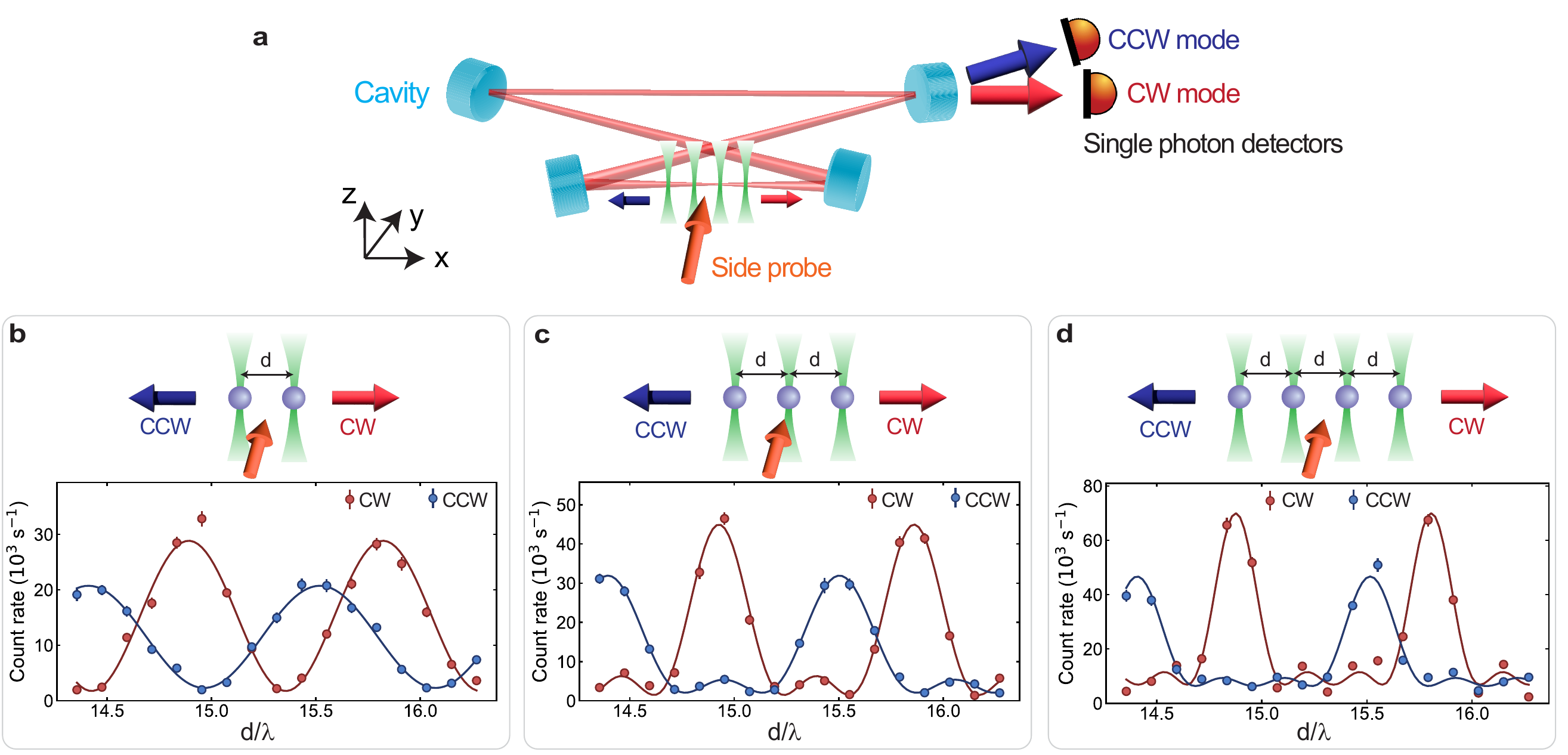} 
    \caption{\textbf{Cavity Bragg scattering.} (a) Optical tweezers, propagating along the $z$-direction, are used to position an array of individual atoms within the mode of a high-cooperativity bow-tie cavity. When probed from the side, atoms scatter light into the CW- and CCW-propagating cavity modes, which are detected by two single photon counting modules.  (b-d) Collective scattering of $N=2,3,4$ atoms as a function of interatomic distance $d$ normalized by the wavelength of the probing light $\lambda$. The data is fit to $S(d)=c_0+c_1|G_{\pm}|^2$, where $G_{\pm}$ is the geometric structure factor for the CW and CCW modes (see text), enabling the precise determination of the average atomic separation via the spatial period of the two-mode interference. A 10~G magnetic field is applied along $z$, $\Delta_{ca}/(2\pi)=-100$~MHz, and three-dimensional cavity cooling is performed prior to collective scattering.} 
    \label{fig:1}
\end{figure*}

\section*{Few-atom Bragg scattering}

The scattering of light by atoms inside an optical cavity is significantly modified in the so-called strong-coupling regime, where light emitted into the cavity, and experienced by the atom, interferes strongly with the driving light. For light resonant with the optical cavity, the scattering into the cavity then exceeds the scattering into all of free space by a factor given by the cavity cooperativity, $\eta$~\cite{VladanClassical}. We study the collective scattering of a one-dimensional array of cesium atoms inside of a high-cooperativity ($\eta=21.0(3)$~\cite{peters_cavity-enabled_2024}) bow-tie ring cavity in which light circulates in a loop, supporting two distinct running-wave cavity modes~\cite{YuTingCavity}. In contrast to a standard two-mirror standing-wave cavity, collective emission of many atoms in a ring cavity depends on the relative separation of atoms, rather than boundary conditions set by the mirrors, due to the translational invariance of the ring geometry~\cite{Courteille_2007,Zimmermann_2018,Hao_PRR_2024}. We trap single atoms within optical tweezers that are equally spaced with a lattice constant $d$ along the cavity mode. As shown in Fig.~\ref{fig:1}(a), the array is illuminated from the side with a probe beam that is tuned near the D2 transition at 852~nm, and atoms can scatter into the clockwise (CW) or counter-clockwise (CCW) cavity modes.

For an array with $N$ atoms, the collective scattering into the cavity is determined by the sum of the fields (amplitude and phase) scattered by each atom. Consequently, the geometric structure of the array plays a significant role in determining the scattering rate and direction.
For an array with two atoms (Fig.~\ref{fig:1}b), collective scattering depends sinusoidally on the atom spacing $d$, while for more than two atoms (Fig.~\ref{fig:1}c-d), constructive interference appears only in a narrow range of atomic separations. The narrowing of the regions of constructive interference and the enhancement in the collective scattering rate with increasing atom number is akin to the interference observed in Bragg scattering, which scales quadratically ($\propto N^2$) with the number of scatterers.

\begin{figure}
    \includegraphics[width=0.40\textwidth]{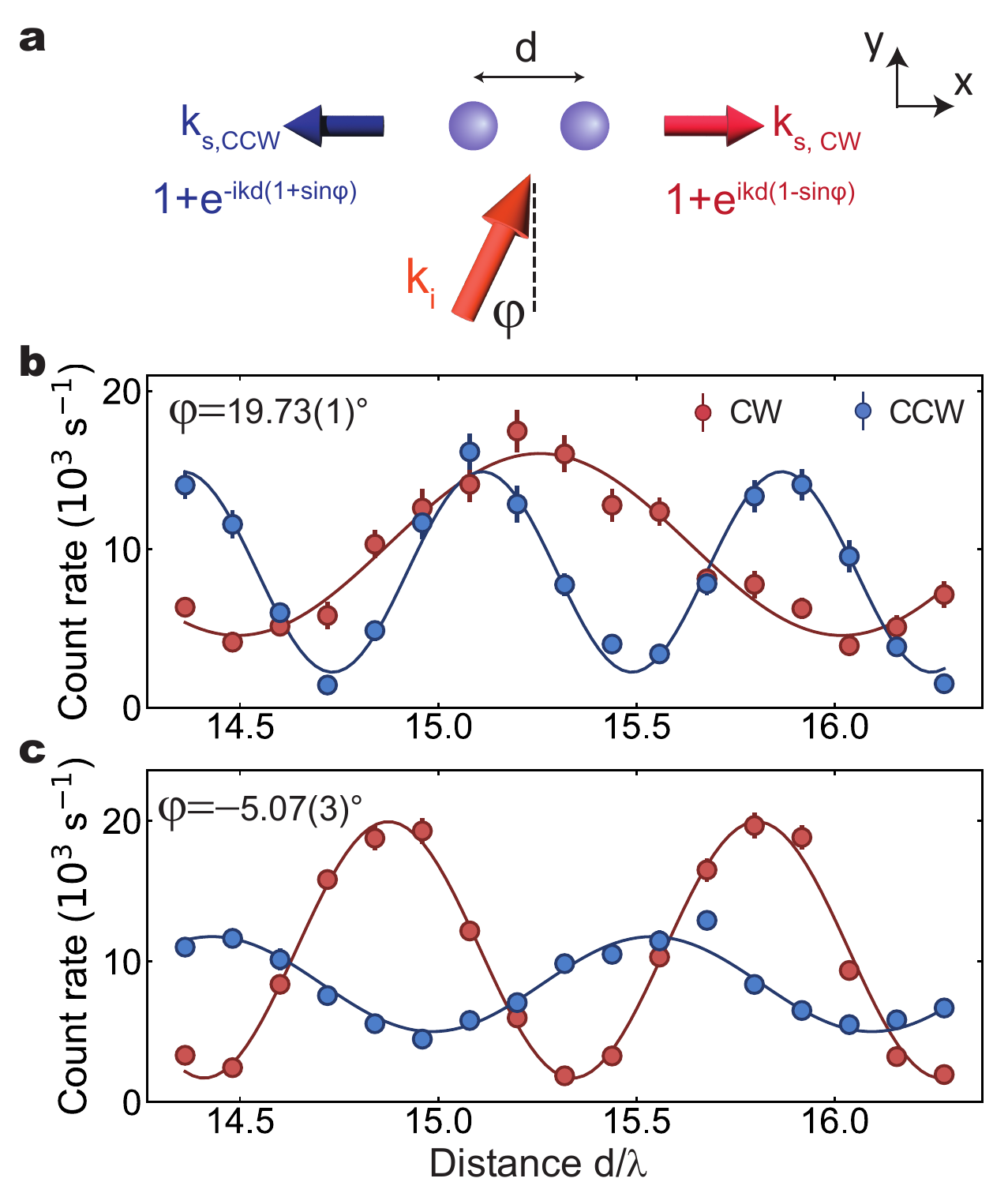} 
    \caption{\textbf{Directional cavity scattering.} (a) Schematic of the geometry leading to asymmetric scattering into the two modes of the bow-tie cavity for nonzero angle of incidence of the probe. (b-c) Collective scattering into the cavity versus interatomic separation for different incident probe angles $\varphi$. A 10~G magnetic field is applied along $z$, and $\Delta_{ca}/(2\pi)=-100$~MHz as before. Here, only cavity cooling of the radial motion is applied. }
    \label{fig:3}
\end{figure}

The condition on the geometric structure to support constructive interference (i.e.~Bragg's condition) for collective scattering into the two-mode bow-tie cavity can be different for the two modes, resulting in asymmetric scattering. The CW/CCW symmetry in scattering is broken by a nonzero angle of illumination $\varphi$ of the probe beam relative to normal incidence to the cavity mode ($y$ in Fig.~\ref{fig:1}a). Specifically, the two geometric factors associated with scattering into the CW- and CCW-propagating modes are 
$G_{\pm}=\sum_n e^{inkd(1  \mp \sin\varphi)}$.
As a result, when varying the lattice constant $d$ the period of interference in the CW and CCW modes is $\lambda/(1\mp \sin\varphi)$.
Near normal incidence, the period of interference in the two modes is similar and Bragg's condition is satisfied at almost the same atomic separation, as shown in Fig.~\ref{fig:1}b-d and Fig.~\ref{fig:3}c. On the other hand, for a large positive incident angle, the interference period for collective scattering into the CW cavity mode becomes much larger than that for the CCW mode, as seen in Fig.~\ref{fig:3}b.

Collective scattering into the two modes provides a precise calibration of the average atom position.
The probe's angle of incidence $\varphi$ can be extracted from the ratio of the interference periods measured in the two counter-propagating modes, $(1+\sin\varphi)/(1-\sin\varphi)$ for the case of two atoms. We fit the experimental data to the form $c_0+c_1|G_{\pm}|^2$, obtaining beam angles of $\varphi=19.73(1)\degree, -5.07(3)\degree$, for the measurements in Figs.~\ref{fig:3}(b-c), respectively.
Provided knowledge of the probe's wavelength $\lambda$, the oscillation period can then be determined with sub-nanometer precision. For example, the periods of oscillation for data collected in the CW and CCW modes in Fig.~\ref{fig:3}(c) are 783.1(4)~nm and 935.0(6)~nm, respectively. 
The same method applies to the case with more atoms in Fig.~\ref{fig:1}(b-d).
Additionally, the precise position measurement provided by the two-mode collective scattering enables calibration of the magnification of the imaging system used for fluorescence imaging. We obtain a value of 6.533(2) averaged across the data in Fig.~\ref{fig:1}(b-d), about 1\% different from the nominal imaging specifications. 

\begin{figure}
    \includegraphics[width=0.495\textwidth]{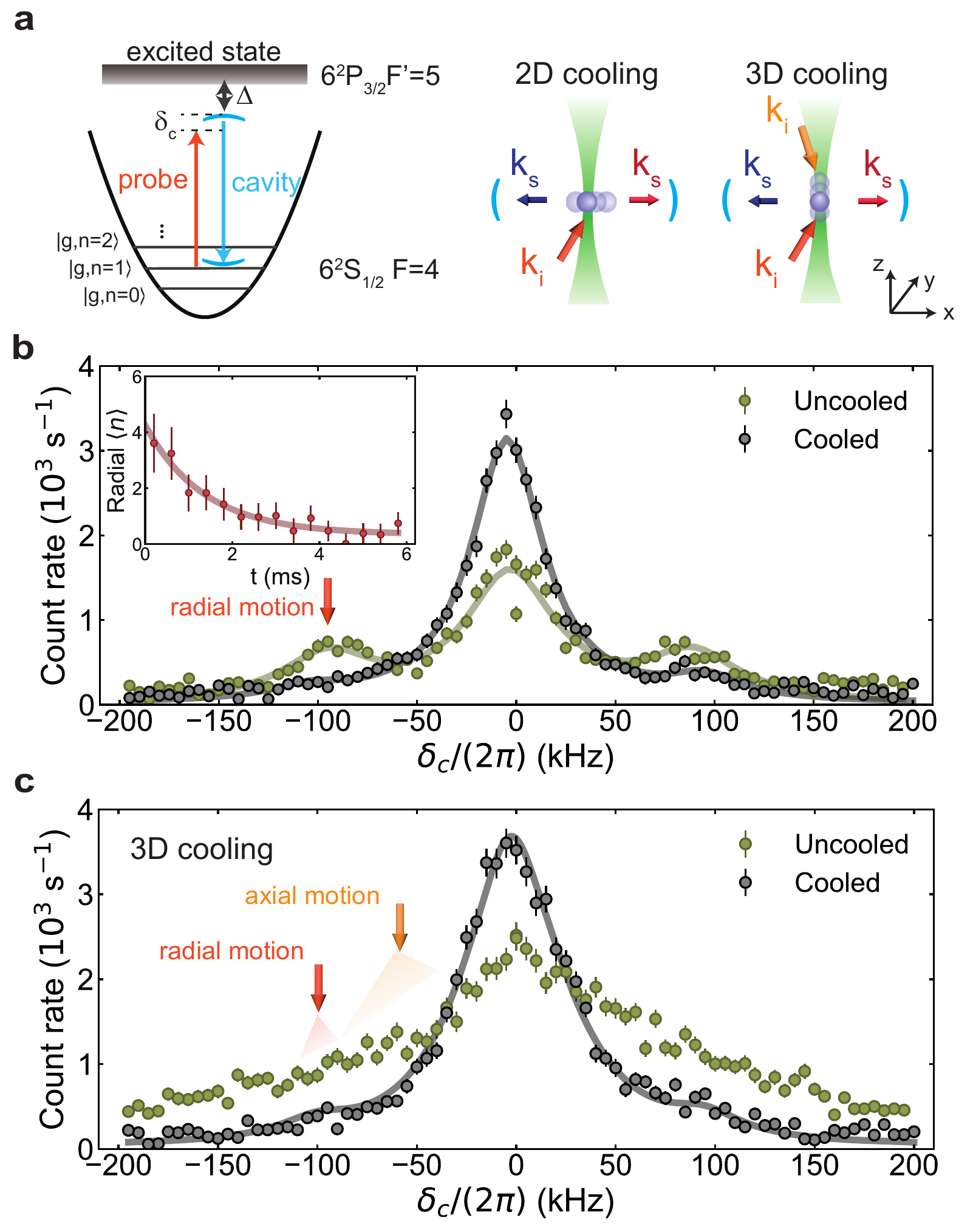} 
    \caption{\textbf{Cavity cooling.} (a) Level diagram for cavity cooling and schematic for cooling of the radial (2D) and both the radial and axial (3D) motions. (b) Sideband spectrum with and without cavity cooling the radial motion of a single atom. 
    A 10~G magnetic field is applied along $z$,  $\Delta_{ca}/(2\pi)=-100$~MHz, and the probe is linearly polarized along the $z$-direction for all data in this figure. The inset of (b) shows the time scale for cavity sideband cooling of $\tau=1.4(3)$~ms. (c) Spectrum with and without cavity cooling of a single atom probed by light with an incident angle of about $45^\circ$ from the $z$-axis. A three-dimensional cooling sequence is achieved by interleaving side and top illumination of the atom. The frequencies for radial and axial cooling are $\delta_c/(2\pi)=-109,-86$~kHz and $\delta_c/(2\pi)=-92,-59,-42,-27$~kHz, respectively, and are marked by the color-coded arrows and shaded regions.}
    \label{fig:2}
\end{figure}
\begin{figure}
\includegraphics[width=0.48\textwidth]{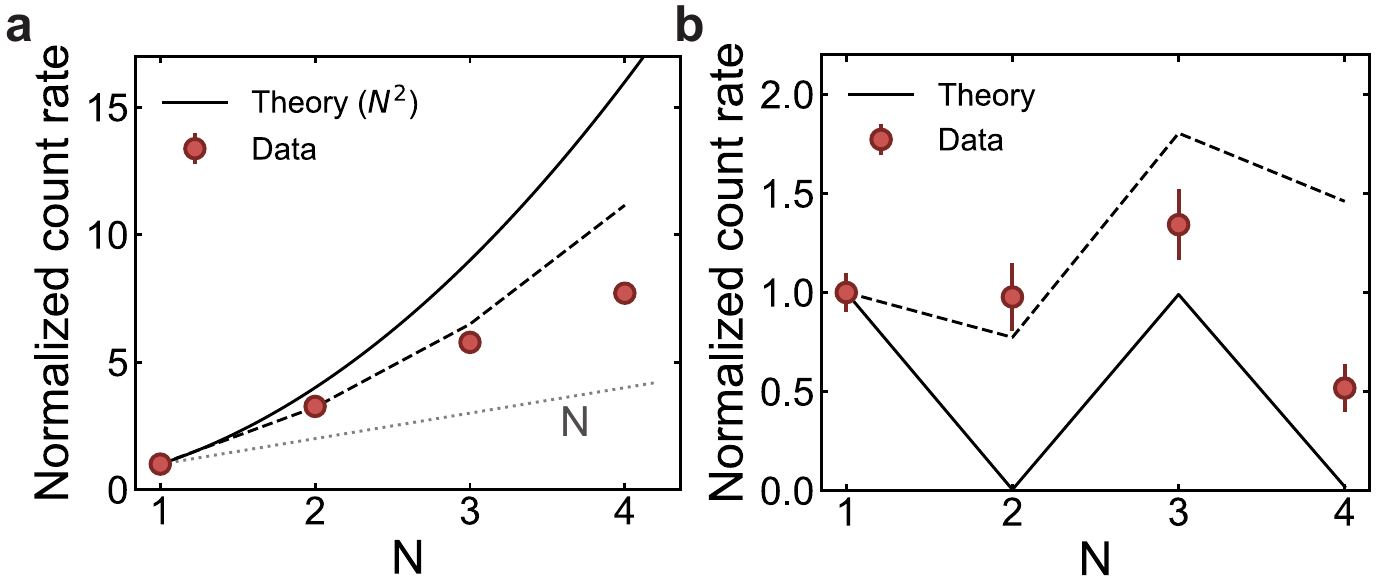} 
\caption{\textbf{Super- and sub-radiant cavity scattering.}  (a) Measurement of super-radiant collective scattering into the CW-propagating cavity mode for $N$ atoms. The tweezer separation is set to $d(1-\sin\varphi)=17\lambda$. (b) Measurement of sub-radiant scattering, with $d(1-\sin\varphi)=17.5\lambda$. For both datasets, a 0.3~G field along $x$ is applied to maximize the scattering into the CW mode and the data is normalized by the single-atom scattering rate. Solid lines show idealized predictions for stationary atoms, while the dashed lines show simulations taking into account coupling to the uncooled axial motion assuming a mean phonon occupation $\langle n_{axial}\rangle=40$ and a Lamb-Dicke parameter $0.17$.}
\label{fig:5}
\end{figure}

\section*{Cavity sideband cooling}
The observations of Bragg scattering are enabled by ground-state cavity cooling. Cavity cooling reduces the vibrational energy of a particle by enhancing the particle's emission on the blue-detuned side of its spectrum, thereby taking away energy during the scattering process~\cite{vuletic_three-dimensional_2001}. 
For free or weakly bound particles, cavity cooling operates in the Doppler limit as the cavity does not spectrally resolve features of individual motional quanta (i.e.~phonons). On the other hand, for tightly trapped particles with sufficient confinement so that transitions between motional states during photon emission are negligible (i.e.~the Lamb-Dicke regime), a cavity with a narrow linewidth can resolve transitions corresponding to addition or removal of single phonons.
Previous studies on coupling neutral atoms in optical cavities have realized cavity Doppler cooling of an atomic ensemble in an optical dipole trap~\cite{hosseini_cavity_2017} and single atoms in standing wave dipole traps~\cite{nusmann_vacuum-stimulated_2005,maunz_cavity_2004}, as well as ground-state cooling with conventional Raman sideband cooling~\cite{boozer_cooling_2006,reiserer_ground-state_2013,urunuela_ground-state_2020}. Here, we utilize our cavity with a linewidth of $\kappa/(2\pi)=37$~kHz to achieve resolved-sideband cavity cooling of individual atoms trapped in optical tweezers.

The motion of a single atom trapped in an optical tweezer includes two radial degrees of freedom (in the $x-y$ plane of Fig.~\ref{fig:2}a) and one axial degree of freedom (along the $z$-direction).
When an atom scatters light from a probe beam incident in the $x-y$ plane into the cavity, the momentum transfer $\vec{k_s}-\vec{k_i}$ can cool both radial degrees of freedom (2D cooling)~\cite{vuletic_three-dimensional_2001}. By scanning the probe frequency and measuring the scattering rate into both modes of the bow-tie cavity, we observe a resolved-sideband spectrum, Fig.~\ref{fig:2}(b), from which we extract a radial trapping frequency of $\omega_r/(2\pi)=89$~kHz. To cool the radial motion, we set the probe-cavity detuning to $\delta_{c}=-\omega_r$ (orange arrow in Fig.~\ref{fig:2}b) such that cavity resonance matches and enhances the blue sideband of the atom's emission spectrum. Each photon scattered into this sideband removes a phonon from the radial motion of the atom. The cavity spectrum of the cooled atom (blue data in Fig.~\ref{fig:2}b) shows a suppressed cooling sideband, as well as an enhanced carrier due to improved atom-cavity coupling at the lower temperature. A separate measurement probing the amplitude of the red and blue sidebands, compensating for the tails of the carrier, after 6~ms of cooling gives a mean phonon occupation of $\langle n_{rad}\rangle=0.17(4)$, corresponding to a temperature of $2.1(3)~\mu$K and a (one-dimensional) ground-state probability of 85\%. Additionally, a time-resolved trace of scattering into the cavity on the cooling sideband indicates a radial cooling timescale of 1.4(3)~ms (inset to Fig.~\ref{fig:2}b).

The axial motion of an atom is cooled by a probe beam sent in the $x-z$ plane such that the momentum transfer couples to motion along the $z$-direction. With the two beams illustrated in Fig.~\ref{fig:2}a, we implement a 3D cooling sequence composed of a radial and several axial cooling pulses applied sequentially, indicated by the arrows in Fig.~\ref{fig:2}(c). Since the axial trapping frequency $\omega_a/(2\pi)\sim20$~kHz is smaller than the cavity linewidth, a broad feature is observed in the uncooled cavity spectrum measured with the axial probe beam. The 3D cooling sequence significantly narrows the measured linewidth, as shown in Fig.~\ref{fig:2}(c), and a fit extracts a phonon occupation of $\langle n_{axial}\rangle=3.4(2.4)$, corresponding to an axial temperature of $3.9(2.4)~\mu$K and a ground-state probability of 23\%.

We further apply cavity cooling to an array of many atoms. 
For multi-atom cooling, the probe beam is detuned by $\Delta_{ca}/(2\pi)=1.521~$GHz to match the resonance of a neighboring longitudinal $\mathrm{TEM}_{00}$ mode of the cavity, corresponding to a different free-spectral range (FSR). At this detuning, the dispersive shift is 0.66~kHz per atom and the linewidth broadening is 2.3~Hz per atom, negligible compared to the cavity linewidth $\kappa/(2\pi)=37$~kHz. We measure a phonon occupation of the radial motion similar to the single-atom case for two, three, and four atoms (see Figure~\ref{fig:temperature}).
When all tweezers have identical trap frequencies, there are collective motional modes that cannot be cooled efficiently due to destructive interference in scattering into the cavity. We find that our tweezers differ in frequency enough to cool each atom in the array to the ground state of its trap. We test this by performing cavity cooling of a multi-atom array, eliminating one of the tweezers, and then observing that the apparent phonon occupation has not changed (see Methods).

We find that performing 3D cavity cooling prior to collective scattering measurements is critical to the high-contrast interference observed in Fig.~\ref{fig:1}. Finite temperature can  smear the geometric structure of the array and increase (incoherent) scattering on the motional sidebands.
In addition, due to imperfect alignment, the probe used for collective scattering has a small coupling to the axial motion of the tweezers. As a result, we find that the axial temperature plays an important role in the collective scattering, as can be seen by comparing Fig.~\ref{fig:3}c, where the axial motion is uncooled (2D cooling), to Fig.~\ref{fig:1}b, where 3D cooling is applied.

We study the scaling of collective scattering with atom number by tuning the inter-atomic distance to points of constructive and destructive interference, $kd(1-\sin\varphi)/(2\pi)=17, 17.5$, respectively, for the CW mode as shown in Fig.~\ref{fig:5}. For constructive interference, an enhanced scattering rate beyond linear scaling is observed (superradiant scattering), while for destructive interference, scattering becomes nearly independent of atom number (subradiant scattering). We attribute the discrepancy from ideal super- and sub-radiant behavior to the finite axial temperature in these datasets, as well as imperfect constructive or destructive interference in scattering from atoms' different Zeeman sublevels exhibiting different Rayleigh scattering rates due to differences in Clebsch-Gordan coefficients~\cite{masson_state-insensitive_2024,yan_superradiant_2023} (see SI for more details).

\section*{Directional light scattering by a single atom}

The data reported so far have been attained under conditions (transverse magnetic field) where the single-atom scattering into the CW and CCW cavity modes is symmetric, and the asymmetry in CW and CCW scattering has derived from the geometric structure of the multi-atom array. We now discuss directional light scattering by a single atom due to the chiral nature of our bow-tie cavity. 

In a four-mirror ring cavity, an out-of-plane twist of the cavity mode, resulting in geometric polarization rotation, easily breaks the degeneracy between modes of a different helicity~\cite{Simon_2018,Simon_Laughlin_2020}. In our system, a frequency splitting of 8.417~MHz is observed between right circularly polarized and left circularly polarized light (relative to the propagation directon), see Fig.~\ref{fig:spectrum}. In the absence of an atom, the $\sigma_+$ ($\sigma_-$) CW-propagating mode is degenerate with the $\sigma_-$ ($\sigma_+$) CCW-propagating mode. Such chiral quantum optics~\cite{lodahl_chiral_2017,Chiral_review_2025} has been studied in nanofibers~\cite{Schneeweiss_chiral_superradiant_burst_2024} and microresonators~\cite{junge_strong_2013,shomroni_all-optical_2014,Rauschenbeutel_2016}. With a magnetic field along the cavity axis, the time-reversal symmetry of the atom-light interaction is broken by an atom prepared in a magnetically sensitive Zeeman state. Accordingly, when probed by light resonant with a cavity mode corresponding to one helicity, the scattering rate into each direction can be very different for a polarized atom.

\begin{figure}
\includegraphics[width=0.48\textwidth]{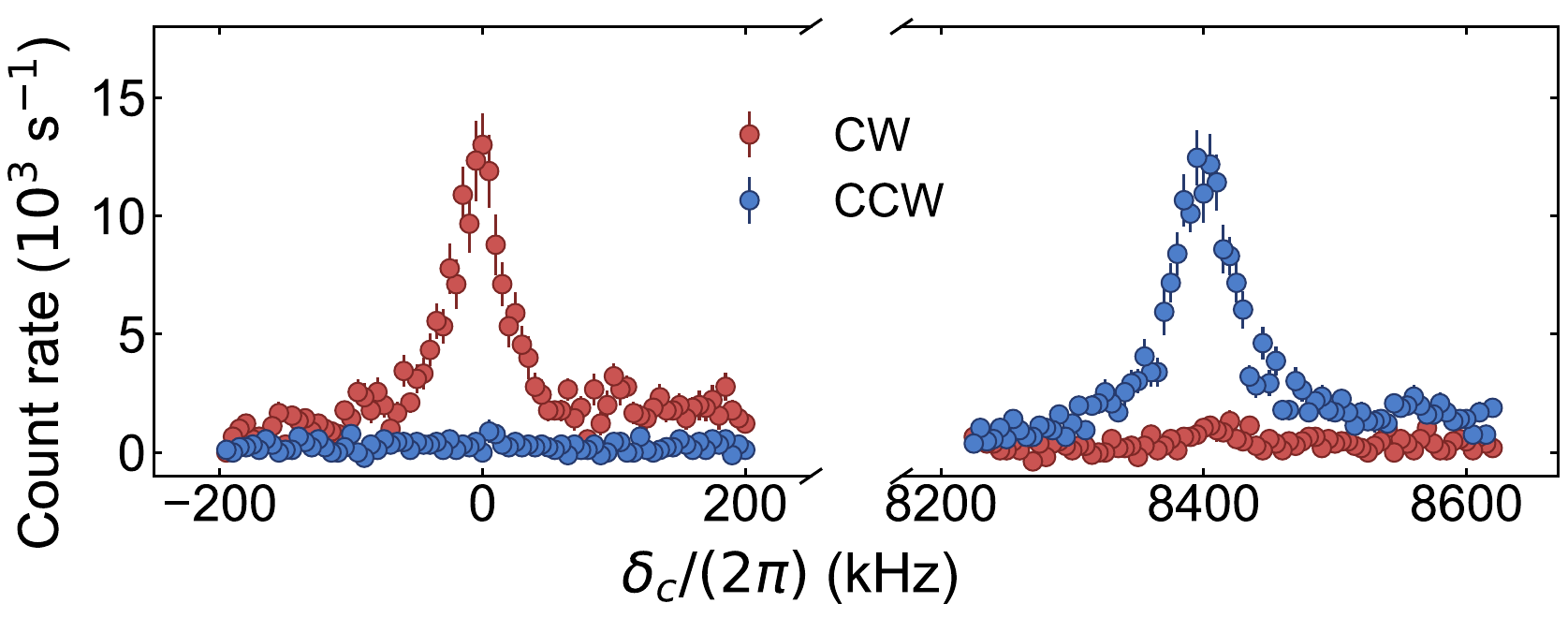} 
\caption{\textbf{Single-atom chiral scattering.} A single atom, prepared in the $\ket{F=4, m_F=+4}$ sublevel, scatters light into the two counter-propagating cavity modes. The driving light is incident normal to the cavity axis, polarized along the $z$-direction, and its frequency is swept across the chiral cavity resonances. Red and blue data points represent light collected in the CW and CCW modes.} 
\label{fig:directional_emission}
\end{figure}

Specifically, we probe a single cesium atom with light near the $F=4$ to $F^\prime=5$ resonance, incident close to orthogonal to the cavity mode, and linearly polarized along the $z$-direction, as before. This light has equal amplitude in the $\sigma_+$/$\sigma_-$ basis defined with respect to the cavity axis. An atom prepared in the $F=4, m_F=+4$ state has coupling strengths to the $\sigma_+$ and $\sigma_-$ components of this light that differ by a factor of 45.
Hence, when sweeping the probe frequency across the resonance corresponding to different helicities of the intra-cavity light, we see strong directional scattering into the cavity by a single atom, as shown in Fig.~\ref{fig:directional_emission}.
Our measurements demonstrate a directionality (ratio of peak scattering rates) of about 23 and 11 for the two cavity resonances. We observe that a magnetic field of about 0.3~G along the cavity axis is sufficient to observe directional emission so long as no large transverse field is applied. We note that for the other measurements presented in this work, we apply a strong magnetic field along the $z$-direction to eliminate the chiral coupling and observe balanced scattering into the two cavity modes for a single atom. 

\section*{Conclusion and outlook}
In summary, we combine the tweezer array platform with a novel bow-tie ring cavity to explore collective light-scattering phenomena at the few-atom level. We observe asymmetric interference in the two running-wave cavity modes governed by the geometric structure of two, three, and four atom arrays. Additionally, we see the narrowing of regions of constructive interference with increasing atom number, matching a Bragg-like condition for scattering even with separations much larger than the optical wavelength. The presence of interference in both cavity modes enables measurement of the average atomic separation with sub-nanometer precision. Furthermore, we present the first realization of ground-state cavity cooling of single atoms trapped in optical tweezers in the resolved-sideband regime, cooling atoms to about 2~$\mu$K along all three directions. Finally, we observe directional light scattering by a single atom in our chiral, twisted ring cavity when the atom breaks the time-reversal symmetry of the light-matter interaction.

Engineering directional light-matter interaction with single-atom chiral light scattering and few-atom Bragg scattering paves the way for studying quantum chiral optics~\cite{Chiral_review_2025,peter_chirality_2024} such as non-Hermitian and non-reciprocal interactions~\cite{rieser_tunable_2022,reisenbauer_non-hermitian_2024}, as well as exploring symmetry-breaking and phase transitions in cavity optomechanics such as self-organization~\cite{ho_optomechanical_2024,domokos_collective_2002,chan_observation_2003,Black_2003,Esslinger_2010,Lev_2018}.
Combined with the fast and non-destructive state detection and atom counting demonstrated recently~\cite{peters_cavity-enabled_2024,grinkemeyer_error-detected_2024,deist_mid-circuit_2022}, our work can also be extended to create molecules from atoms and study atomic collisions in real-time. In contrast to conventional cooling methods, cavity cooling does not depend on the detailed atomic structure, and applies to a broad class of particles including neutral atoms, ions~\cite{leibrandt_cavity_2009}, molecules~\cite{lev_prospects_2008}, and nanoparticles~\cite{delic_cavity_2019,windey_cavity-based_2019,delic_cooling_2020}. In view of the very low attainable temperatures given by the cavity linewidth (2 $\mu$K in our case), cavity cooling can be used to improve the fidelity of various quantum processes, such as the gate fidelity in Rydberg arrays~\cite{evered_high-fidelity_2023}.

\acknowledgements
This work is supported by the U.S. Department of Energy, Office of Science, National Quantum Information Science Research Centers, Quantum Systems Accelerator. Additional support is acknowledged from the NSF Frontier Center for Ultracold Atoms (grant number PHY- 2317134), the NSF QLCI Q-SEnSE (grant number QLCI-2016244), 
and the DARPA ONISQ programme (grant number 134371-5113608).
U.D. acknowledges the support of the Austrian Science Fund (FWF) 10.55776/STA175.

\section*{Methods}
\textbf{Modes of bow-tie cavity.}
The schematic of our setup is shown in Fig.~\ref{fig:1}a. A one-dimensional array of tweezer-trapped cesium atoms is placed at the small waist of a bow-tie cavity along the $x$ direction~\cite{YuTingCavity}.  The traps are generated by magic-wavelength 937~nm light~\cite{Magic_Kim_2003, Magic_Ye_1999}  using an out-of-vacuum microscope objective (NA=0.5) and an acousto-optic deflector. The waist of each trap is 1~$\mu$m with a trap depth $U/h=27$~MHz. The large spacing (3.1~cm) between the cavity mirrors of the short arm enables direct loading of the atom array from a magneto-optical trap (MOT) generated inside the cavity. Two 852~nm beams near the cesium D2 transition are sent from the side in the $x-y$ and $x-z$ planes to probe the scattering into the cavity by the atoms.

\begin{figure}[h]
    \includegraphics[width=0.49\textwidth]{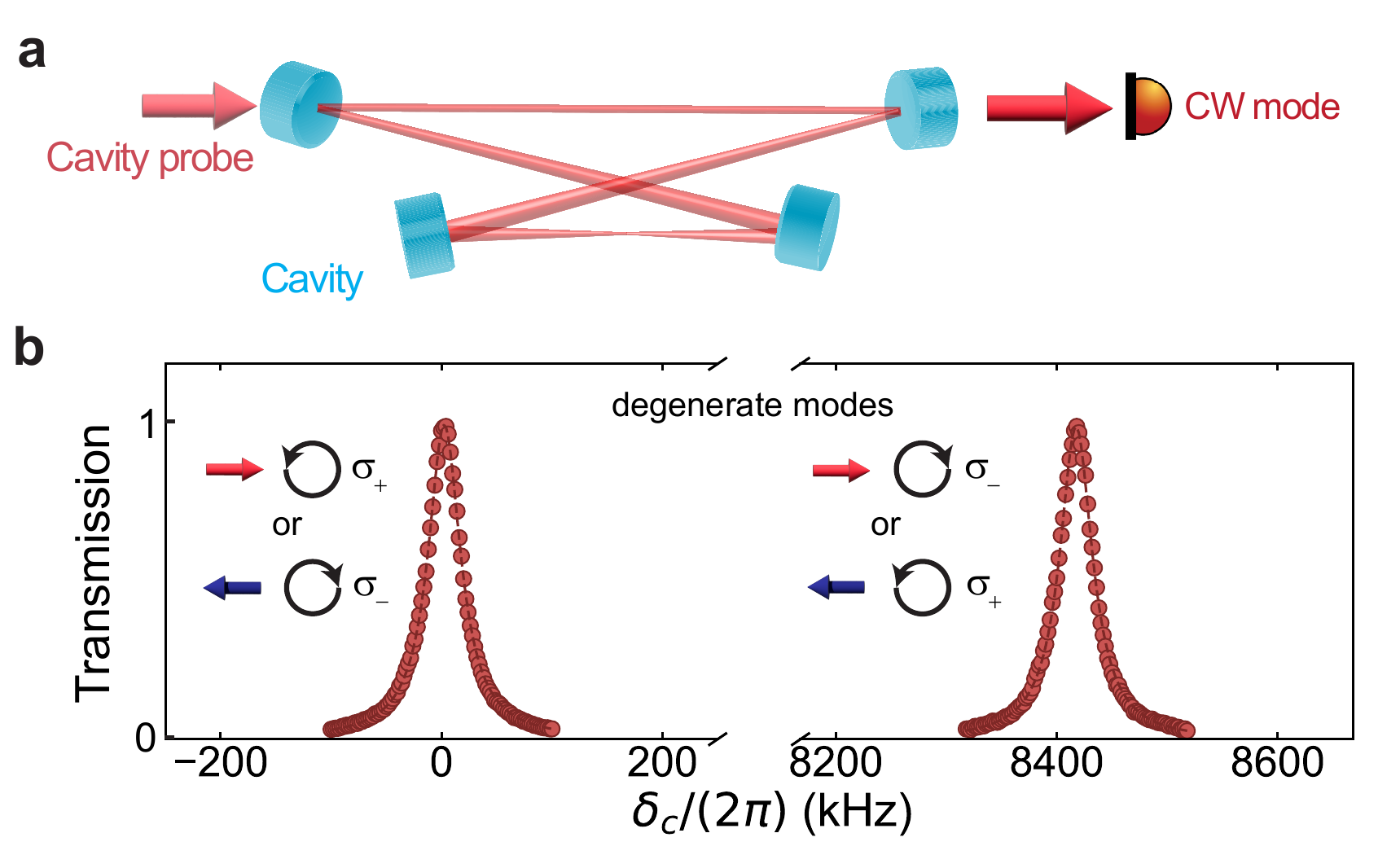} 
    \caption{\textbf{Empty-cavity spectrum.} (a) Schematic for probing the empty-cavity spectrum. Light is incident on one of the cavity mirrors to probe the CW mode, and transmission is collected on a single photon counting module. (b) Measured empty-cavity spectrum.}
    \label{fig:spectrum}
\end{figure}

We denote the two counter-propagating running-wave modes in our cavity as clockwise (CW) and counterclockwise (CCW) modes. A small out-of-plane twist of the cavity breaks the degeneracy of $\sigma_+$ and $\sigma_-$ polarizations (defined relative to the $x$ axis) for a given direction, yielding a frequency splitting of 8.417~MHz as shown in Fig.~\ref{fig:spectrum}b. Due to time-reversal symmetry, the frequency of the $\sigma_+$ ($\sigma_-$) CW mode is degenerate with the $\sigma_-$ ($\sigma_+$) CCW mode in the absence of an atom in the cavity. The measured linewidth and cooperativity of a single propagating mode are $\kappa/(2\pi)=36.7$~kHz and $\eta=4g^2/\kappa\Gamma=21$, respectively~\cite{peters_cavity-enabled_2024}.

\begin{figure}[h]
    \includegraphics[width=0.495\textwidth]{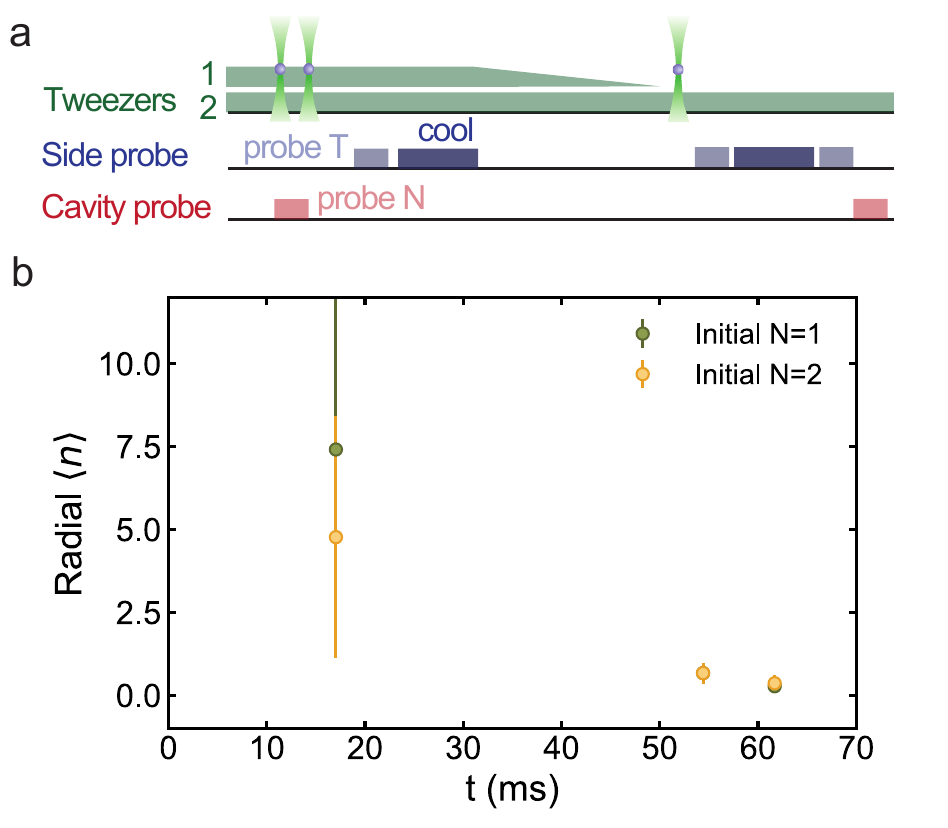} 
    \caption{\textbf{Phonon occupation of one atom in an array.} (a) Schematic of the sequence to test coupling to collective motion. A two-atom array is prepared via probabilistic loading and measurement of atom number via cavity transmission near-atomic resonance~\cite{peters_cavity-enabled_2024}. A side probe cavity cools the array and is also used to measure the amplitude of the resolved-sideband spectrum in order to extract the mean phonon occupation. This probe is detuned from atomic resonance by $\Delta_{ca}/(2\pi)=1.706$~GHz. (b) Extracted phonon number during the sequence: before cavity cooling, after cavity cooling and ramping down of one tweezer, and after cooling the remaining atom.}
    \label{supp_fig:ramp}
\end{figure}

\begin{figure}
    \includegraphics[width=0.35\textwidth]{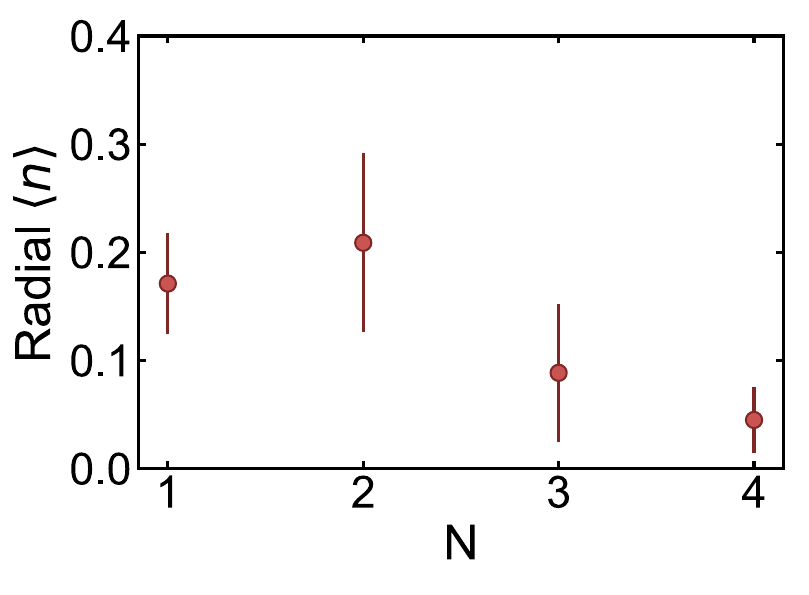} 
    \caption{\textbf{Radial phonon number of cavity cooled multi-atom arrays.} Mean phonon occupation of the radial motion of arrays with different atom numbers. The probe is detuned from atomic resonance by $\Delta_{ca}/(2\pi)=1.521$~GHz.}
    \label{fig:temperature}
\end{figure}

\textbf{Cavity cooling of a multi-atom array.}
To test the coupling of the cavity cooling to collective motional modes, we measure the phonon occupation of a single atom after applying cavity cooling to a two-atom array as shown schematically in Fig.~\ref{supp_fig:ramp}a. 
We begin by preparing a two-atom array probabilistically, and identify instances with two atoms by probing cavity transmission~\cite{peters_cavity-enabled_2024}. We cavity cool the radial motion of the array as in the main text with a beam incident from the side of the cavity and detuned by $\Delta_{ca}/(2\pi)=1.706$~GHz. We measure the mean phonon occupation in the radial motion by probing the carrier and sidebands of the resolved-sideband spectrum before cavity cooling as well as after cavity cooling and adiabatically ramp down one of the tweezers so that only one atom remains. Finally, we cool the remaining atom again and measure its temperature to test whether it can be cooled further. Additionally, we perform the same procedure but with only one atom initially loaded in the preserved tweezer, in order to distinguish any heating due to coupling between tweezer tones in the AOD. We observe no measurable difference between the phonon occupation of a single atom when it is cooled in the presence of another atom or by itself, indicating little coupling to collective motion while cavity cooling a multi-atom array. We observe a small additional temperature reduction after the final cooling stage of the remaining atom in both two- and one-atom cases, perhaps indicating heating during the ramping down of the other tweezer trap. Separately, we cavity cool multi-atom arrays with different atom numbers and observe radial phonon occupations demonstrating cooling to the ground state, Fig.~\ref{fig:temperature}.

We also measure the heating rate of the tweezer traps by cavity cooling, waiting for a variable time, and then measuring the radial phonon occupation. This measurement is limited by the single atom lifetime in the traps, which derives from the background pressure in our vacuum chamber. We bound the heating rate to be below $10\text{ phonon}/$s in our experiment, while the anticipated heating rate due to photon scattering of the far-detuned tweezer light is on the order of 1 phonon/second.

\newpage

\bibliography{main_text} 

\onecolumngrid
\clearpage
\begin{center}
\textbf{\large Supplemental Material: Cavity-enabled real-time observation of individual atomic collisions}
\end{center}
\setcounter{section}{0}
\setcounter{equation}{0}
\setcounter{figure}{0}
\setcounter{table}{0}
\setcounter{page}{1}
\makeatletter
\renewcommand{\theequation}{S\arabic{equation}}
\renewcommand{\thesection}{S\arabic{section}}
\renewcommand{\thefigure}{S\arabic{figure}}

   
\section{Light scattering of an array of two-level atoms in a ring cavity}
The Hamiltonian of an array of atoms at position $\{r_j\}$ coupled to a ring cavity under a tilted side drive is given by
\begin{equation}
    H =-\Delta_c(a_r^\dagger a_r+a_l^\dagger a_l)-\sum_j\Delta_a\sigma_j^\dagger\sigma_j +g\sum_j (\sigma_j^\dagger a_r e^{ikr_j} +\sigma_j^\dagger a_l e^{-ikr_j}+h.c.) + \frac\Omega2\sum_j (\sigma_j^\dagger e^{ikr_j\sin\varphi} +h.c.).
    \label{eq:H_multiatom}
\end{equation}
Here $a_{r,l}$ are the annihilation operators of the two traveling modes, $\Delta_c=\omega_p-\omega_c$ is the detuning between drive and cavity, and $\Delta_a=\omega_p-\omega_a$ is the detuning between the drive and atomic resonance. $2g$ is defined as the Rabi frequency of a single photon in the right (or left) mode flopping the atomic level, $\Omega$ is the Rabi frequency of the side drive, and $\sigma_j$ are the atomic lowering operators. In the low saturation limit, solving for the steady-state atomic excitation and inserting it into Eq.~\eqref{eq:H_multiatom} yields the effective Hamiltonian
\begin{align}
    H=&-\Delta_c (a_r^\dagger a_r + a_l^\dagger a_l) +\sum_j \frac{g^2\Delta_a}{(\frac{\Gamma}{2})^2+\Delta_a^2}\left(a_r^\dagger a_r +a_l^\dagger a_l + e^{-2ikr_j}a_r^\dagger a_l + e^{2i kr_j}a_r a_l^\dagger \right)\\
    &+\sum_j \frac{g\Omega\Delta_a}{2[(\frac{\Gamma}{2})^2+\Delta_a^2]}\left(e^{-ikr_j (1-\sin\varphi)}a_r^\dagger + e^{ikr_j(1+\sin\varphi)}a_l^\dagger+e^{ikr_j (1-\sin\varphi)}a_r + e^{-ikr_j(1+\sin\varphi)}a_l\right)\\
    &=-(\Delta_c -NU_e)(a_r^\dagger a_r + a_l^\dagger a_l) + U_e(G_0 a_r^\dagger a_l  +G_0^* a_r a_l^\dagger) + \Omega_e (g a_r^\dagger + G_l a_l^\dagger +h.c.).
    \label{eq:Heff_multiatom}
\end{align}
where in the last line we have defined $U_e =\frac{g^2\Delta_a}{(\frac{\Gamma}{2})^2+\Delta_a^2}$, the effective drive of the cavity field $\Omega_e = \frac{2g\Omega\Delta_a}{\Gamma^2+4\Delta_a^2}$, and three geometric factors $G_0=\sum_j e^{-2ikr_j}$, $G_r=\sum_j e^{ikr_j (1-\sin\varphi )}$, and $G_l=\sum_j e^{-ikr_j (1+\sin\varphi)}$. To calculate the steady-state photon numbers in right and left modes, we can solve the Heisenberg equations of motion for the field operators in each mode. In the far-detuned limit such that the cross-coupling between the right and left modes can be neglected, the photon numbers in the two modes are
\begin{align}
    n_r&=|G_r|^2 \frac{g^2\Omega^2}{\Gamma^2+4\Delta_a^2}\frac{4}{\kappa^2[1+N\eta \mathcal{L}(\Delta_a)]^2+4(\Delta_c-NU_e)^2},\\
    n_l&=|G_l|^2 \frac{g^2\Omega^2}{\Gamma^2+4\Delta_a^2}\frac{4}{\kappa^2[1+N\eta \mathcal{L}(\Delta_a)]^2+4(\Delta_c-NU_e)^2}.
\end{align}

\section{Cavity Cooling Temperature Measurements}

In the resolved-sideband regime, where the cavity linewidth is much smaller than the trap frequency, $\kappa\ll\omega_t$, the mean phonon occupation, $\langle n\rangle$, can be experimentally measured by probing the cavity scattering rate at carrier and both sidebands. In the Lamb-Dicke regime and for temperatures approaching the groundstate the mean phonon occupation can be calculated from
\begin{equation}
    \langle n\rangle = \frac{\Gamma_{red}}{\Gamma_{blue}-\Gamma_{red}}\label{eq:radial_phonon}.
\end{equation}
On the other hand, when the mean phonon occupation of the system is not close to 0 and the sidebands are not resolved, one must take into account a distribution of sidebands with higher order components and with weights given by a thermal distribution in order to extract a temperature from the spectrum~\cite{wineland_laser_1979}. Assuming the system is under thermal equilibrium with a mean phonon occupation $\langle n\rangle = 1/(e^{\hbar \omega_t/k_B T}-1)$ according to a Bose-Einstein distribution, then the occupation $p_l$ on each level $\ket{l}$ is given by 
\begin{equation}
    p_l=\frac{e^{-l\frac{\hbar\omega_t}{k_BT}}}{Z}= \exp(-l\frac{\hbar\omega_t}{k_BT})\left[1-\exp(-\frac{\hbar\omega_t}{k_BT})\right].
\end{equation}
The scattering rate into the $m^{th}$ sideband $\omega_p+m\omega_t$ should then be calculated with
\begin{align}
    \Gamma_{m,T} &=\Gamma_w\times P(m),\ \text{with }\\
    P(m) &= \sum_{n_l=-m}^{\infty} p_l |\bra{n_l+m}e^{i(k_i-k_s)x_t}\ket{n_l}|^2\nonumber\\&= \exp[-\frac{1}{2}m\frac{\hbar\omega_t}{k_B T} -  (2\langle n\rangle +1)\eta_{LD}^2]\times I_m\left[\exp(\frac{\hbar\omega_t}{2k_B T})2\langle n\rangle\eta_{LD}^2 \right]
    \label{eq:Pm}
\end{align}
where $I_m(\alpha)$ is the modified Bessel function of the first kind, and $P(m)$ is the scattering fraction of the $m^{th}$ sideband~\cite{wineland_laser_1979}. When probing the scattered light using a cavity, one gets the spectrum
\begin{equation}
    \Gamma_c(\omega_p)=\eta\Gamma_{sc}\sum_{m=-\infty}^{\infty}\frac{\kappa^2}{\kappa^2+4(\omega_p+m\omega_t-\omega_c)^2}\times P(m)\label{eq:unresolved_spectrum}.
\end{equation}
Thus, scattering from the $m^{th}$ red(blue) sideband of the measured cavity spectrum at $\omega_p=\omega_c\mp m\omega_t$ has a weighted amplitude of $P(\pm m)$. When a light scattering process couples to multiple vibrational degrees of freedom, the spectrum contains multiple manifolds of sidebands. Assuming the motions are uncoupled, the resulting spectrum is a simple extension of Eq.~\ref{eq:unresolved_spectrum}.

In our experiment, we extract the radial temperature after cavity cooling through a separate measurement of the red and blue sidebands. We apply cavity cooling for 6~ms before each measurement of the scattering rate at the carrier, red, and blue sideband frequencies. We ensure that the phonon number is not modified during these measurements by setting the probe beam intensity such that average scattered photon number is less than 1. With knowledge of the measured cavity linewidth, we then subtract off the contribution of the tails of the carrier transition at the frequencies of the red and blue sidebands in using Eq.~\ref{eq:radial_phonon} to extract the average phonon number. The temperature can then be calculated from $\exp(\frac{\hbar\omega}{k_BT}) = (1+\langle n\rangle)/\langle n\rangle$.

\begin{figure*}
    \includegraphics[width=0.75\textwidth]{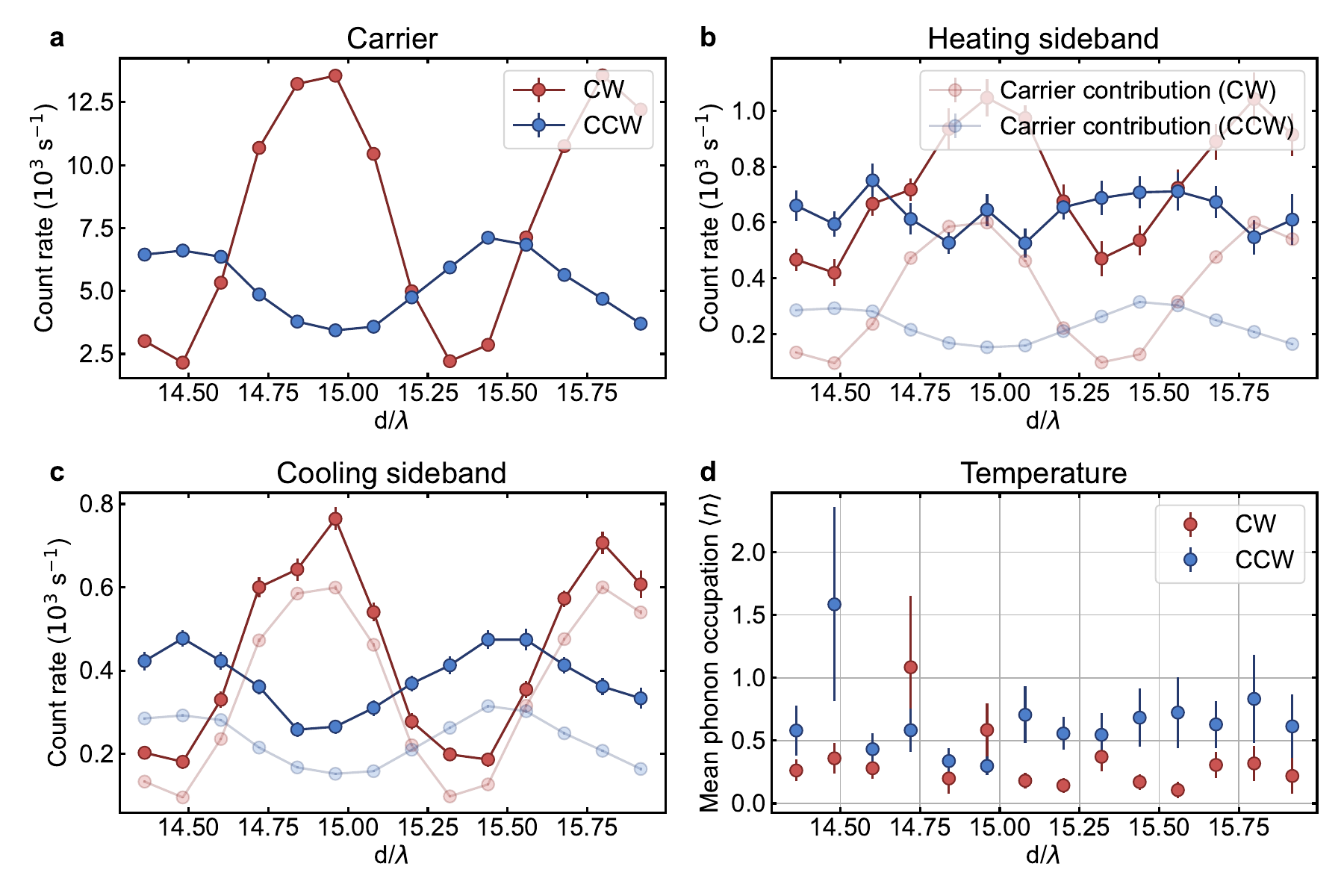} 
    \caption{{\bf Phonon occupation of a two-atom array.} (a) Dependence of the collective scattering rate of two atoms on their atomic separation when probed (a) on cavity resonance (i.e.~the carrier transition), (b) on the blue sideband of radial motion, and (c) on the red sideband of radial motion. The light red and blue markers show the inferred contribution from carrier-scattering, assuming a cavity linewidth of $\kappa=(2\pi)36.7$~kHz. (d) The extracted mean phonon occupation using the measurements in (b) and (c). Here, the cavity-atom detuning is $\Delta_{ca}=1.521$~GHz and cavity cooling of the radial motion is applied for 6~ms before either measurement of the carrier scattering rate for 1~ms, or the sidebands for 0.2~ms. A magnetic field of about $10$G is applied transverse to the cavity.}
    \label{supp_fig:temp_2atom}
\end{figure*}
Due to the collective interference in cavity scattering on the carrier transition, the temperature measurement is more complicated for a multi-atom array. One must subtract off the carrier contribution at the sideband frequencies for the geometry used while cavity cooling. To validate that our temperature measurement is consistent regardless of the geometric structure of the atom array, we measure the temperature of a two-atom array for various atomic separations $d/\lambda$, Fig.~\ref{supp_fig:temp_2atom}. We observe that the modulation of the collective scattering into the cavity at all probe frequencies is accounted for by Rayleigh scattering from the carrier transition. Accordingly, the extracted mean phonon number, Fig.~\ref{supp_fig:temp_2atom}d, is independent of atomic separation to our measurement resolution. We note that the temperature measured by the left mode is higher than the temperature measured by the right mode. This results from coupling to axial motion in scattering into the left mode due to imperfect alignment of the tweezers, probe, and cavity mode. For the measurements in Fig.~\ref{supp_fig:temp_2atom}, we only apply radial cavity sideband cooling, while the axial motion remains hot. The same effect reduces the contrast in Bragg scattering as discussed in the main text.

To extract the temperature of the axial motion, we directly fit the measured spectrum (main text Fig.~5c) using Eq.~\eqref{eq:unresolved_spectrum} with both axial and radial couplings. We use a six parameter fit for the cavity resonant frequency $\omega_c$, the two trapping frequencies $\omega_{axial},\omega_{radial}$, mean phonon occupation $\langle n_{axial}\rangle,\langle n_{radial}\rangle$, and cavity scattering rate $(\eta\Gamma_{sc})$. We set the Lamb-Dicke parameter of the two motions to be $\eta_{axial}=\sqrt{E_{rec}/\hbar\omega_{axial}}$, $\eta_{radial}=\sqrt{E_{rec}/\hbar\omega_{radial}}$ where $E_{rec}=(2\pi)2.07$~kHz assuming that the beam is incident near the longitudinal axis of the tweezer and is orthogonal to the cavity mode. To simply the fitting process, we truncate the order of the included sideband in the cavity spectrum to $|m|\leq 4$, $|n|\leq 2$ and $mn=0$ for the axial and radial sidebands, respectively.

\subsection{Radial Temperature Limit Considerations}
For a Lamb-Dicke trap, cavity-cooling in the resolved-sideband regime has a cooling limit given by~\cite{vuletic_three-dimensional_2001}
\begin{equation}
    \langle n\rangle = \left(\frac{\kappa^2}{\kappa^2+16\omega_t^2}D\eta+ C \right)\frac{1}{D\eta (1-\frac{\kappa^2}{\kappa^2+16\omega_t^2})}\approx\frac{\kappa^2}{16\omega_t^2} + \frac{1}{\eta}\frac{C}{D}(1+\frac{\kappa^2}{16\omega_t^2}),
\end{equation} 
where $D$ is the ratio of the recoil energy of the cavity scattering to the free-space recoil energy $E_{rec,0}=\hbar^2k^2/2m$, and $C$ is the ratio set by the direction of the beam and dipole orientation. Both constants $D$ and $C$ are on the order of unity.
In our experiment, the radial motion of the tweezer is in the $x-y$ plane with a trapping frequency $\omega_t\sim(2\pi)90$~kHz, yielding $\kappa^2/(16\omega_t^2)\approx0.01$. Considering cavity scattering into both $+x$ and $-x$ directions of the bow-tie cavity (with equal rate) to cool the atomic motion along the $x$ direction gives $D_x=2$, $C_x=2/5$, 
while for the atomic motion along the $y$ direction yields $D_y=2$, $C_y=7/5$, 
and therefore a cooling limit corresponding to an average radial phonon occupation of $\langle n_{rad}\rangle\approx 9/(20\eta)$. The single-atom, single-mode cooperativity of our cavity is $\eta\approx21$ when probing near-resonance of the $F=4,m_F=+4\leftrightarrow F=5,m_F=+5$ cycling transition that couples optimally to the circularly polarized cavity mode. When probing the atom from the side of the cavity, along $y$, with a large detuning compared to the excited state energy structure, the contribution from different excited states as well as the splitting of the polarization modes reduces the effective single-mode cooperativity to $\frac23\times\frac12\eta=7$. Thus, we estimate the cooling limit of the radial motion in our system to be $\langle n_{rad}\rangle\approx 0.07$.

\section{Incoherent collective-scattering from multi-level atoms}
For atoms with more than two levels involved in the light scattering process, entanglement between the scattered light and the atoms' state can introduce incoherence in collective light scattering as measurement of the photon field effectively traces over the states of the atoms~\cite{chan_quantum_2003}. For Cesium atoms, we consider the transitions between the $F=4$ ground states and  $F^\prime=5$ excited states of the Cesium D2 line. We assume the initial state of the atom is pumped to a mixed state and coupled to a single mode of the cavity in vacuum
\begin{equation}
\rho_0=\sum_{m=-4}^{m=+4}c_m\ket{m}\bra{m}\otimes\ket{0}\bra{0},
\end{equation}
where the $c_m$ indicate the distribution of population over the Zeeman sublevels, satisfying $\sum c_m=1$. Consider a Rayleigh scattering process such that the population distribution over Zeeman sublevels is preserved, the atom-photon state of the system after the scattering process becomes
\begin{equation}
\rho=\sum_{m=-4}^{m=+4}c_m\ket{m}\bra{m}\otimes\ket{\alpha_m}\bra{\alpha_m},
\end{equation}
where $\ket{\alpha_m}$ is a coherent state with field amplitude $|\alpha_m|$ and photon number $|\alpha_m|^2$. Scattering from each Zeeman sublevel occurs with different strengths due to differences in Clebsch-Gordan coefficients, leading to distinct $\alpha_m$ for scattering from each transition in the most general case. 
For two atoms, the scattered field additionally has a relative phase $\phi$ and the state after the scattering process can be written
\begin{equation}
    \rho=\sum_{m=-4}^{m=+4}\sum_{l=-4}^{l=+4}c_m c_l\ket{m}\bra{m}\otimes\ket{l}\bra{l}\otimes\ket{\alpha_m+\alpha_l e^{i\phi}}\bra{\alpha_m+\alpha_l e^{i\phi}}.
\end{equation}
The average scattered photon number is then 
\begin{align}
\langle n\rangle = \Tr{a^\dagger a \rho} &= \sum_{m,l} c_m c_l |\alpha_m+\alpha_l e^{i\phi}|^2 \\&= 2\sum_{m} c_m |\alpha_m|^2+\sum_{m,l}c_m c_l (\alpha_m \alpha_l^* e^{-i\phi}+\alpha_m^* \alpha_l e^{i\phi})\\
&=2\sum_{m} c_m |\alpha_m|^2 + 2\bigg|\sum_{m,l}c_m c_l \alpha_m^* \alpha_l\bigg|\cos{(\phi+\phi_0)},
\end{align}
where $\phi_0 = \arg{(\sum c_m c_l \alpha^* \alpha_l)}$.
When sweeping the phase $\phi$ (e.g.~through the separation between atoms), the collective scattering into the cavity thus oscillates with a contrast given by
\begin{equation}
    C=\frac{|\sum c_m c_l \alpha_m^* \alpha_l|}{\sum c_m |\alpha_m|^2} = \frac{\sum c_m \alpha_m^* \sum c_l \alpha_l}{\sum c_n |\alpha_m|^2}=\frac{|\sum c_m \alpha_m|^2}{\sum c_m |\alpha_m|^2},
\end{equation}
which is less than 1 by a Cauchy-Schwarz inequality.

\begin{figure}
    \centering
    \includegraphics[width=0.5\linewidth]{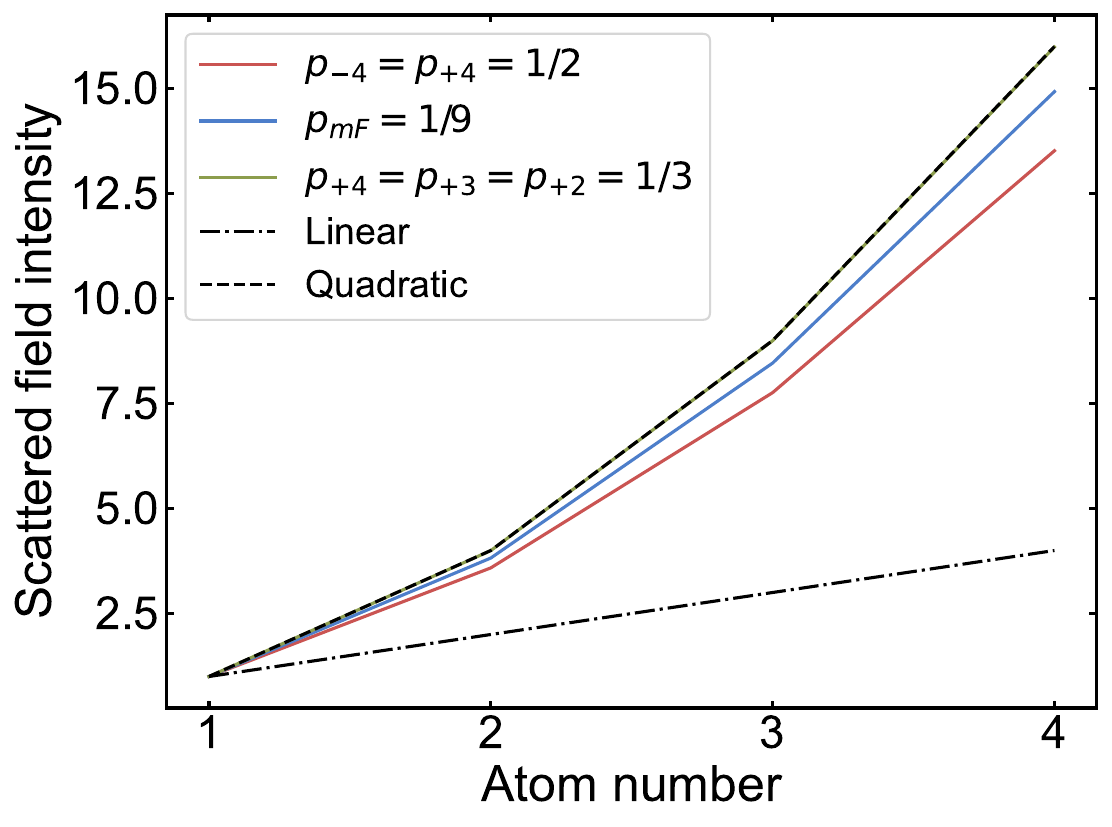}
    \caption{\textbf{Collective-scattering at Bragg's condition for multi-level atoms.} Cavity-atom detuning is set to $\Delta_{ca}/(2\pi)=100$~MHz. We assume the probe beam has a linear polarization, orthogonal to the quantization axis of the atoms. Each curve represents the collective scattering from atoms with different initial populations in Zeeman sublevels indicated by the legend.
    }
    \label{supp_fig:atom-photon-entanglement}
\end{figure}

In order to understand the role multi-level scattering plays in collective-scattering in our system, we estimate the scattering for arrays of atoms matching the Bragg condition for a few representative initial states in Fig.~\ref{supp_fig:atom-photon-entanglement}, with an atomic detuning of $\Delta_{ca}/(2\pi)=100$~MHz in order to make the trend apparent. In the most extreme case when the initial state has half the population in each of the stretched states, the effective incoherence from tracing over light-atom entanglement has the most dramatic effect. In the case of pumping to near a stretched state $p_{+2}=p_{+3}=p_{+4}$, the interference is very close to $N^2$ scaling, while the result for equal population in all sublevels falls in between these cases. When studying this scaling experimentally, we remove the strong transverse magnetic field so that emission is directional, into the CW mode, and the atom is pumped at least close to the $F= 4, m_F=4$ state. Additionally, the experimental detuning is $\Delta_{ca}/(2\pi)=1.521$~GHz, where the scattering rate from different hyperfine levels will be more similar than in the example of Fig.~\ref{supp_fig:atom-photon-entanglement}. Hence, we conclude that the incoherence due to scattering from many sublevels is a small effect in our experiment and instead we believe it is likely the finite temperature, perhaps induced by the directional scattering, that is the largest contributor to the measurements showing less than $N^2$ scaling.

\section{Spectrum of Collective Scattering}
\begin{figure*}[h]
    \includegraphics[width=0.95\textwidth]{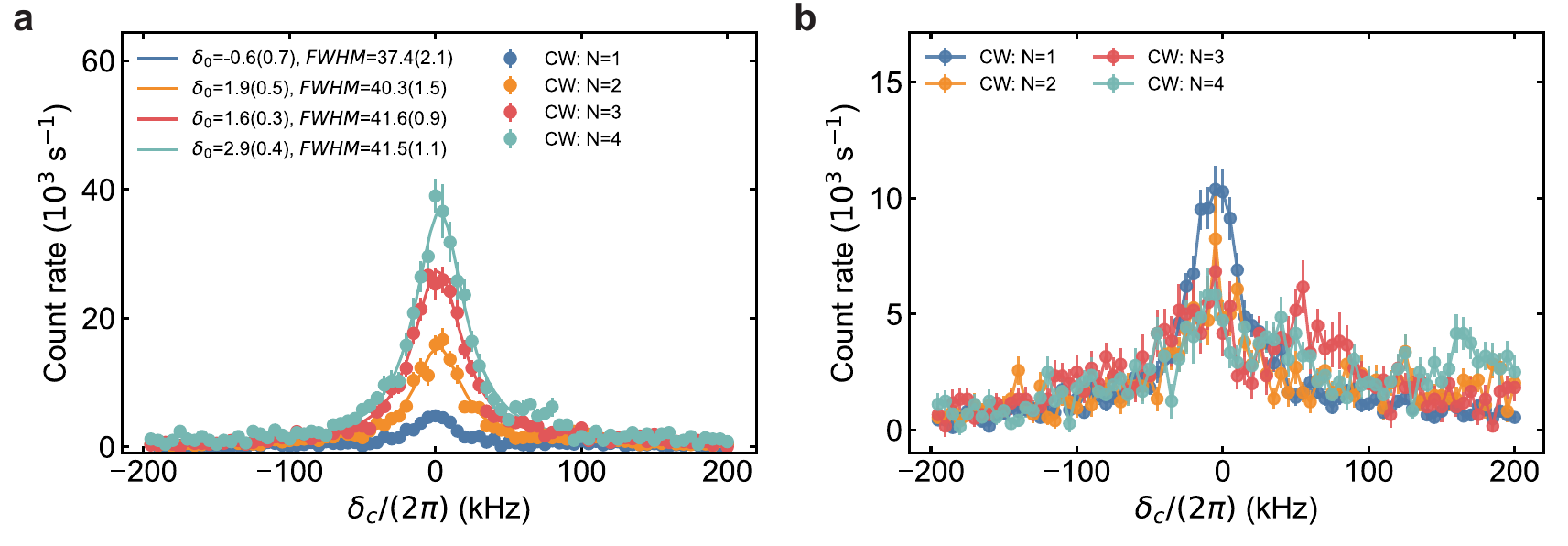}
    \caption{{\bf Spectra for collective cavity scattering.} (a),(b) Spectra for collective scattering by $N=1,2,3,4$ atoms separated with a distance $d(1-\sin\varphi)/\lambda=17,17.5$, respectively. The cavity-atom detuning is $\Delta_{ca}=1.521$~GHz. In these measurements, we apply cavity cooling to cool the radial motion.}
    \label{supp_fig:supsubrad}
\end{figure*}

In Fig.~\ref{supp_fig:supsubrad} we show the spectrum from collective scattering measured at atomic separations that yield constructive and destructive interference. The data in main text Fig.~4 are extracted from Fig.~\ref{supp_fig:supsubrad}a,b by calculating the mean scattering rate in the range of $[-10,10]$~kHz. We note that in these measurements we apply a small magnetic field along the cavity axis to maximize the cavity scattering into the CW mode. As the probing light is tuned far from atomic resonance, the cavity resonance and linewidth are insensitive to the atom number.

\clearpage

\end{document}